\documentclass[preprint,prd,aps,showpacs,showkeys,nofootinbib]{revtex4}

\usepackage[dvips]{graphicx}

\usepackage{dcolumn}

\usepackage{bm}
\usepackage{epsfig}
\usepackage{psfrag}

\def\be{\begin{equation}}

\def\ee{\end{equation}}

\begin{document}

\title{CP-phase effects on the effective neutrino mass $m_{ee}$ in the case of quasi-degenerate neutrinos}

\author{J. Maalampi$^{a,b}$ and J. Riittinen$^{a}$}

\affiliation{$^a$ Department of Physics, P.O.~Box~35, FIN-40014 University of Jyv\"askyl\"a, Finland}
\affiliation{$^b$ Helsinki Institute of Physics, P.O.~Box~64, FIN-00014 University of Helsinki}

\date{\today}

\bigskip

\begin{abstract}
We study the possibility that the three mass states of the ordinary active neutrinos actually split into pairs of  quasi-degenerate  states, with $\Delta m^2_{kk'} \sim 10^{-12}$ eV$^2$ or less, as a result of mixing of active neutrinos with sterile neutrinos. Although these quasi-degenerate pairs will look in laboratory experiment identical to  single active states, the CP phase factors associated with active-sterile mixing might cause cancellations in  the effective electron neutrino mass $m_{ee}$ measured in the neutrinoless double beta decay experiments thereby revealing the split nature of states. 
\end{abstract}

\keywords{Neutrino mixing, sterile neutrino, CP violation}

\maketitle

The possible existence of active-sterile neutrino mixing with a high mass degeneracy  will not be revealed  in oscillation experiments probing atmospheric, solar or laboratory produced neutrinos
if the degeneracy is of the order of $\Delta m^2 \sim 10^{-12}$ eV$^2$ or less \cite{Berezinsky}. However, as we have pointed out earlier, one might obtain hints of such degeneracy through the effects it may have on the measured fluxes of the standard active neutrinos originating in Active Galagtic Nuclei (AGN)\cite{UHECR} or in supernovae \cite{supernova1,supernova2} (for a recent study, see e.g. \cite{Esmaili}). For example, the ratio $\nu_e$ : $\nu_{\mu}$ : $\nu_{\tau} = 1:1:1$ of neutrino fluxes from AGN's one would expect in the standard three-neutrino-scheme \cite{Bento} could be substantially distorted by the mixing of the active neutrinos with their quasi-degenerate sterile counterparts.  

In this paper we will examine another possible indication of the existence of quasi-degenerate neutrino pairs, namely the effects these pairs may have, through the extra CP-phases associated with their mixing, on the so called effective electron neutrino mass $m_{ee}$ probed in neutrinoless double beta decay ($(\beta\beta)_{0\nu}$) experiments.  The neutrinoless double beta decay, which breaks the conservation of lepton number by two units, is particularly interesting process as it is the only practical way to reveal the possible Majorana nature of neutrinos. Furthermore, the process is at the moment  the only viable way to  obtain information about the absolute mass scale of light neutrinos. 

Generally, the effective mass $m_{ee}$ is given by
\begin{equation}\label{eq1}
m_{ee} = \left| \sum_{i}U^2_{ei}m_i \right|,
\end{equation}
where the sum runs over all the mass states $\nu_i$ that has a $\nu_e$ component, $U$ is the neutrino mixing matrix ($\nu_{\ell}=\sum_{i} U_{\ell i}\nu_i$), and $m_i$ is the mass of the state $\nu_i$.
The value of  $m_{ee}$ depends on the phase factors of the matrix $U$, and the cancellations between different terms of the sum are possible. As a result of the cancellations, $m_{ee}$ could be smaller than any of the  $m_i$. The effect of CP-phases on the $m_{ee}$ has been previously studied in the standard three-neutrino case eg. in \cite{3nu} and in the four-neutrino case in \cite{4nu}. 

We will consider a scenario where there exist, in addition to the three active neutrino flavours $\nu_e, \nu_{\mu}, \nu_{\tau}$, three sterile neutrino flavours $\nu_{s1}, \nu_{s2}, \nu_{s3}$. In the standard three-neutrino case the three active neutrinos mix which each other and form three mass eigenstates. We will denote these states as $\hat{\nu}_k$ ($k=1,2,3$). In our model these states are not  mass eigenstates but we assume each of them to  mix with a sterile state and to form with it two orthogonal superpositions which are quasi-degenerate in mass. In the experiments  not sensitive enough to recognize the mass difference of the states such quasi-degenerate states would look like a single state, the corresponding active state $\hat{\nu}_k$. 

The mass
eigenstates that result from the mixing of the active state $\hat{\nu}_k$ and  the sterile state $\nu_{sk}$  are the following:
\begin{eqnarray}\label{states}
\nu_k & = & \cos{\varphi_k} \hat{\nu}_k - e^{i \delta_k} \sin{\varphi_k} \nu_{sk} \\
\nu_{k'} & = & e^{i \delta_k} \sin{\varphi_k} \hat{\nu}_k + \cos{\varphi_k} \nu_{sk}.
\end{eqnarray}
The CP-phases $\delta_k$ can have the values $0$ or $\pi$ if CP is conserved, in the CP violating case the phase angle has some other value.  The mass squared difference within each pair is assumed to be
$\Delta m^2_{kk'} \sim 10^{-12}$ eV$^2$, and the mass differences between different pairs are the same as those of the mass states in the standard three neutrino model.

The matrix elements $m_{\alpha \beta}$ between two neutrino flavours has the following general expression in terms of the neutrino masses  \cite{Whisnant}:
\begin{equation}
m_{\alpha \beta} = \left| \sum_{k=1}^{3} \tilde{U}^*_{\alpha k} \tilde{U}^*_{\beta k} m_k + \sum_{k'=1}^{3} \tilde{U}^*_{\alpha k'} \tilde{U}^*_{\beta k'} m_{k'} \right|,
\end{equation}
where $m _k$, $m_{k'}$ are the masses of the neutrino mass eigenstates and $\tilde{U}_{\alpha k}$ the elements of the unitary $6\times 6$ mixing matrix $\tilde{U}$
which diagonalizes the $6 \times 6$ neutrino mass matrix.
When considering the quantity $m_{ee}$ only the matrix elements of first row of $\tilde{U}$ are relevant. They  are given by
\begin{eqnarray*}
\tilde{U}_e &=& \left(
\begin{array}{cccccc}
\tilde{U}_{e1} & \tilde{U}_{e2} & \tilde{U}_{e3} & \tilde{U}_{e1'} & \tilde{U}_{e2'} & \tilde{U}_{e3'} \\
\end{array} \right) \\
&=& \left(
\begin{array}{cccccc}
\cos{\varphi_1}U_{e1}, & \cos{\varphi_2}U_{e2}, & \cos{\varphi_3}U_{e3}, & e^{-i \delta_1}\sin{\varphi_1}U_{e1}, &  e^{-i \delta_2}\sin{\varphi_2}U_{e2}, & e^{-i \delta_3}\sin{\varphi_3}U_{e3} \\
\end{array}
\right),
\end{eqnarray*}
where $U_{ek}$ are elements of the standard $3\times 3$ neutrino mixing matrix as defined in \cite{UHECR}.

Let us denote the mass difference between eigenstates $\nu_k$ and $\nu_{k'}$ as the  $\epsilon_k$, that is, $m_{k'} = m_k + \epsilon_k$. We will then have
\begin{equation}
m_{ee} = \left| \sum_{k=1}^{3} \left( \cos^2{\varphi_k} + e^{2i\delta_k} \sin^2{\varphi_k} \right) U^2_{ek} m_k + \sum_{k=1}^{3} e^{2i\delta_k} \sin^2{\varphi_k} U^2_{ek} \epsilon_k \right|.
\label{m_ee}
\end{equation}
As the mass difference $\epsilon_k$ is according to our assuption very small, we can  approximate it as
\begin{equation}
\epsilon_k \approx \frac{\Delta m^2_{kk'}}{2 m_k}.
\end{equation}

Generally, there may be CP-phases in the standard $3 \times 3$ active neutrino mixing matrix $U$ also. Let us assume that this is not the case here but that all  CP-phases arise from the mixings
of the active and sterile states. The values of the mixing angles and the squared mass differences are most stringently constrained by cosmology. If they are too large, neutrino oscillations would bring sterile neutrinos into thermal equilibrium with other particles leading to a conflict between theoretical and observational result concerning the primordial nucleosynthesis \cite{cosmo}. Because the mass differences between degenerate states are in our case very small, this constraint will not cause any problem and the active sterile mixing angles $\varphi_k$  can have any value between $0$ and $\pi / 4$. 

The CP-phases associated with the active-sterile mixing may cause cancellations in $m_{ee}$. As one can see, a value of $\delta_k$ different from $0$ causes 
cancellation in the first terms of eq. (\ref{m_ee}), and with $\delta_k = \pi / 2$ the cancellation is complete if the active-sterile mixing is maximal, i.e. $\varphi_k = \pi / 4$. In this situation the states $\nu_k$ and $\nu'_k$ would combine into Dirac particle, providing their mass difference $\epsilon_k$ vanishes, and the neutrinoless double beta decay would be prohibited. If the mass difference does not vanish, the decay is still possible but the decay width would be tiny, dictated by the mass difference.
The effect of the mixing of the states $3$ and $3'$ is very small due to fact that these states contain only a tiny  $\nu_e$ component. In the case of the lighter states,
on the other hand, the effect is noticeable, and can decrease the value of $m_{ee}$ by by a factor of about $2/3$ and $1/3$ for the $1$-$1'$ and $2$-$2'$ -mixings, respectively. 

Considering the absolute mass scale of neutrinos, the most stringent limit comes from cosmology. According to the  result of the WMAP collaboration, the sum of  masses of the light active neutrinos is constrained as   \cite{WMAP}
\begin{equation}
\sum_k m_k < 0.67 \, \text{eV}.
\label{masslimit}
\end{equation}
This also gives the absolute upper limit for the effective mass $m_{ee}$ in the limit of exact degeneracy, as one can infer from Eq. (\ref{eq1}).
Let us assume that this limit is saturated and that the values of the squared mass differences are small in comparison with the mass limit (\ref{masslimit}) we make an approximation that all the mass states are, more or less, degenerate. 
Then
\begin{equation}
m_k \approx 0.2 \, \text{eV},
\end{equation}
where $k = 1,...,6$.

\begin{figure}[h!]
\centering
\includegraphics[height=10cm]{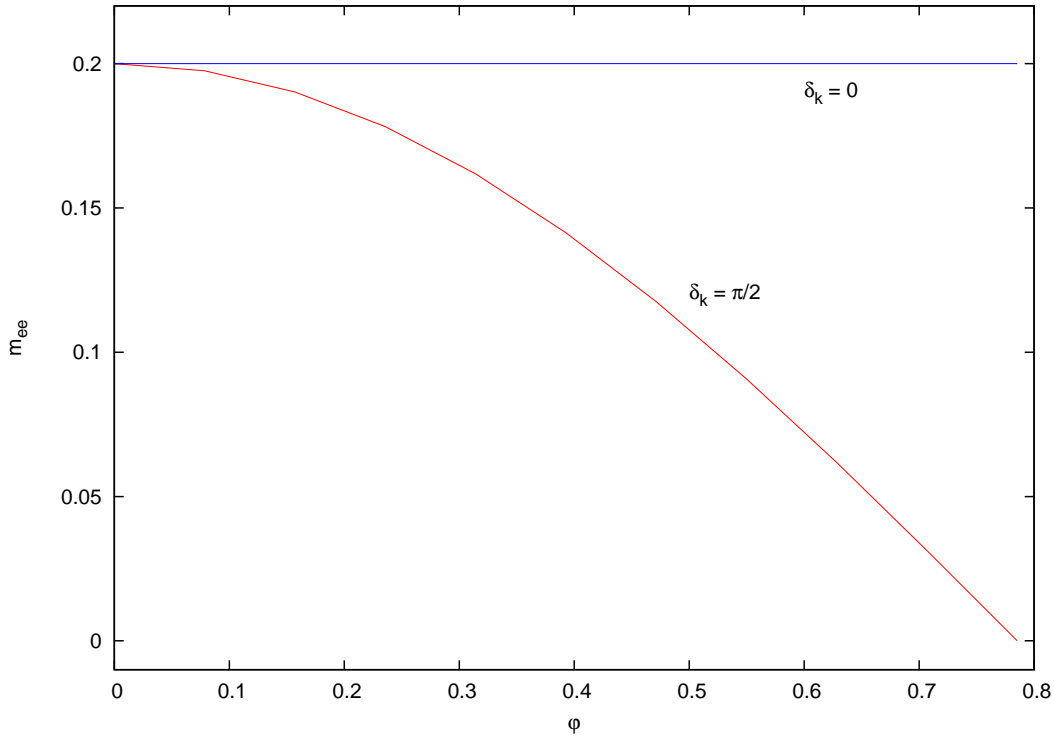}
\caption{\small The effect of all three active-sterile mixings on the value of $m_{ee}$. All the active-sterile mixing angles have the same value $\varphi_k = \varphi, \, k=1,2,3$.}
\label{fig1}
\end{figure}

Using these numbers, we plot in Fig. \ref{fig1} the value of $m_{ee}$ as a function of the active-sterile mixing angle $\varphi$.  As can be seen from the plot, the combined effect of the 
all three mixings can cause a strong cancellation, decreasing the value of $m_{ee}$ almost to zero at large mixing angles. We remind that here we have ignored the possible cancellations arising from the phases appearing in the standard $3\times 3$ mixing matrix. They could make the cancellation even stronger.

\

In conclusion, possible extra CP-phases, which come from the existence of extra neutrino states, could greatly affect the effective neutrino mass $m_{ee}$ via cancellations. With suitable CP-phases even quite small mixings between active and sterile neutrino states can decrease the value of $m_{ee}$ notably. This would affect
the analyses of and conclusions made from  the results of neutrinoless double beta decay experiment, which probe the value of $m_{ee}$.  The measurement of $m_{ee}$ is currently the only way to get information about absolute scale of neutrino masses. It is important to keep in mind this new possible source of cancellation when drawing conclusions from the data.  



\begin{thebibliography}{99}
\bibitem{Berezinsky} V. Berezinsky, M. Narayan and F. Vissani, Nucl. Phys. B 658, 254 (2003).
\bibitem{UHECR} P. Ker\"anen, J. Maalampi, M. Myyryl\"ainen and J. Riittinen, Phys. Lett. B 574, 162 (2003).
\bibitem{supernova1} P. Ker\"anen, J. Maalampi, M. Myyryl\"ainen and J. Riittinen, Phys. Lett. B 597, 374 (2004).
\bibitem{supernova2} P. Ker\"anen, J. Maalampi, M. Myyryl\"ainen and J. Riittinen, Phys. Rev. D 76, 125026 (2007).
\bibitem{Esmaili} A. Esmaili, arXiv:0909.5410 [hep-ph].
\bibitem{Bento}
  J.~G.~Learned and S.~Pakvasa,
  Astropart.\ Phys.\  {\bf 3}, 267 (1995);
  L.~Bento, P.~Keranen and J.~Maalampi,
  Phys.\ Lett.\  B {\bf 476}, 205 (2000);
  H.~Athar, M.~Jezabek and O.~Yasuda,
  Phys.\ Rev.\  D {\bf 62}, 103007 (2000).
\bibitem{3nu} H. Minakata and O. Yasuda, Phys. Rev. D 56, 1692 (1997); T. Fukuyama, K. Matsuda and H. Nishiura, Phys. Rev. D 57, 5844 (1998); R. Adhikari and G. Rajasekaran, Phys. Rev. D 61, 031301(R) (1999);
V. Barger and K. Whisnant, Phys. Lett. B 456, 194 (1999); M.Czakon, J. Gluza and M. Zralek, Phys.Lett. B 465, 211 (1999).
\bibitem{4nu} A. Kalliom\"aki and J. Maalampi, Phys. Lett. B 484, 64 (2000); S. M. Bilenky, S. Pascoli and S. T. Petcov, Phys. Rev. D 64, 113003 (2001); S. Pakvasa and P. Roy, Phys. Lett. B 535, 181
(2002).
\bibitem{Whisnant} V. Barger, Yuan-Ben Dai, K. Whisnant and Bing-Lin Young, Phys. Rev. D 59, 113010 (1999).
\bibitem{cosmo}  
K.~Enqvist, K.~Kainulainen and J.~Maalampi,
  Phys.\ Lett.\  B {\bf 249}, 531 (1990);   
  K.~Kainulainen,
  Phys.\ Lett.\  B {\bf 244}, 191 (1990);
  R.~Barbieri and A.~Dolgov,
  Phys.\ Lett.\  B {\bf 237}, 440 (1990);  
  K.~Enqvist, K.~Kainulainen and M.~J.~Thomson,
  Nucl.\ Phys.\  B {\bf 373}, 498 (1992).
\bibitem{WMAP} E. Komatsu et al. (WMAP Collaboration), Astrophys. J. Suppl. 180, 330 (2009), arXiv:0803.0547.
\end{thebibliography}
\end{document}